\documentclass[11pt,a4paper]{article}

\usepackage[hyperref]{emnlp2020}
\usepackage{times}
\usepackage{latexsym}

\usepackage{microtype}

\usepackage[ruled]{algorithm2e} % For algorithms
\usepackage{booktabs}
\usepackage{footnote}
\usepackage{courier}
\usepackage{url}
\usepackage{graphicx}
\usepackage{color}
\usepackage{array}
\usepackage{subfigure}
\usepackage[T1]{fontenc}
\usepackage{aecompl}
\usepackage{float}
\usepackage{amssymb}
\usepackage[mathscr]{eucal}

\newcommand{\ie}{\emph{i.e.}\xspace} % that is
\newcommand{\eg}{\emph{e.g.}\xspace} % for example
\newcommand{\nop}[1]{}
\usepackage[font=small,labelfont=bf,textfont=md]{caption}

\aclfinalcopy % Uncomment this line for the final submission
\def\aclpaperid{1457} %  Enter the acl Paper ID here

\makeatletter
\newcommand{\printfnsymbol}[1]{%
  \textsuperscript{\@fnsymbol{#1}}%
}
\makeatother
\newcommand{\para}[1]{\smallskip\noindent\textbf{#1}}

\title{Will This Idea Spread Beyond Academia? Understanding Knowledge Transfer of Scientific Concepts across Text Corpora}
\author{Hancheng Cao\thanks{Equal contribution}\\
   Stanford University \\
   \texttt{hanchcao@stanford.edu} \\\And
   Mengjie Cheng\printfnsymbol{1} \\
   Harvard Business School \\
   \texttt{macheng@hbs.edu} \\\And
      Zhepeng Cen\printfnsymbol{1} \\
   Carnegie Mellon University \\
   \texttt{zcen@andrew.cmu.edu} \\\AND
      Daniel A. McFarland \\
   Stanford University \\
   \texttt{dmcfarla@stanford.edu} \\\And
    Xiang Ren \\
   University of Southern California \\
   \texttt{xiangren@usc.edu} \\}

\begin{document}

\maketitle
\begin{abstract}

What kind of basic research ideas are more likely to get applied in practice?
There is a long line of research investigating patterns of knowledge transfer, but it generally focuses on \textit{documents} as the unit of analysis and follow their transfer into practice for a specific scientific domain. Here we study translational research at the level of \textit{scientific concepts} for \textit{all scientific fields}. We do this through text mining and predictive modeling using three corpora: 38.6 million paper abstracts, 4 million patent documents, and 0.28 million clinical trials. We extract scientific concepts (i.e., phrases) from corpora as instantiations of ``research ideas", create concept-level features as motivated by literature, and then follow the trajectories of over 450,000 new concepts (emerged from 1995-2014) to identify factors that lead only a small proportion of these ideas to be used in inventions and drug trials.
Results from our analysis suggest several mechanisms that distinguish which scientific concept will be adopted in practice, and which will not. We also demonstrate that our derived features can be used to explain and predict knowledge transfer with high accuracy. Our work provides \textit{greater understanding of knowledge transfer} for researchers, practitioners, and government agencies interested in encouraging translational research. 
\end{abstract}
\begin{figure}[t]
    % \centering
    \vspace{-0.4cm}
    \hspace{-0.7cm}
    \includegraphics[width=1.07\linewidth]{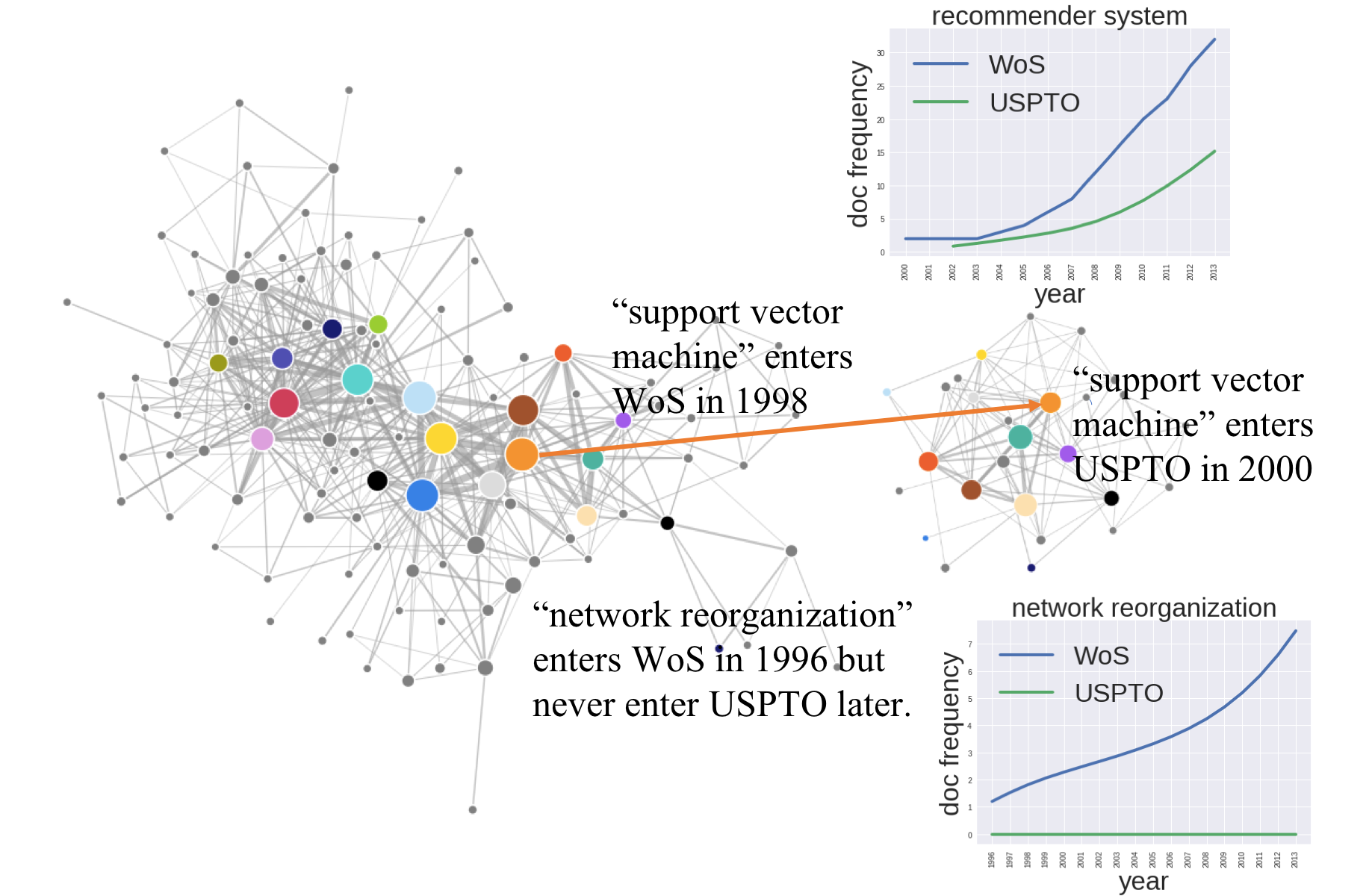}
    \vspace{-0.1cm}
    \caption{
    \textbf{An illustration of scientific concept's ``knowledge transfer" from basic research to practice use}: we analyze individual concept's time-varying features (e.g., popularity) and relative positions with other concepts (i.e., co-occurrence) to understand the key mechanisms behind knowledge transfer, using Web of Science research papers, USPTO patents and clinical trial documents.
    % \textbf{An illustration of scientific concept's \textit{knowledge transfer} from basic research to practice:}
    }
    \label{fig:translation}
    \vspace{-0.3cm}
\end{figure}

\section{Introduction}
Science generates a myriad of new ideas, only some of which find value in practical uses~\cite{backer1991knowledge,lane2011measuring}.
Large government agencies (\eg, NSF, NIH) pour billions of dollars into basic research in the hopes that it will span the research-practice divide so as to generate private sector advances in technologies~\cite{narin1985technology}, social policies~\cite{mcdonald2010social}, and pharmaceuticals~\cite{berwick2003disseminating}. To this end, these agencies increasingly seek to nurture ``translational research" that succeeds at extending, bridging and transforming basic research so it finds greater applied value~\cite{li2017applied}. Surrounding this effort has arisen a line of research that tries to identify when, where, and how academic research influences science and technological invention~\cite{backer1991knowledge, li2017applied}. 

However, prior research efforts are limited in their ability to understand and facilitate the translation of research ideas. This is partially due to a shortage of data, a biased focus on successful examples, and specialized modeling paradigms. 
In practice, only a small proportion of knowledge outputs are successfully translated into inventive outputs ($\sim2.7\%$ concepts from WoS to patent, and $\sim11.3\%$ concepts from WoS to clinical trials, according to our data analysis). 
Previous studies conduct post-hoc analyses of successful scientific-technological linkages, but are unable to explain why the majority of scientific innovations do not transfer into technological inventions. 
Additionally, prior work mostly look at document-level linkages across research and applied domains, i.e., citations from patents into research papers or shared inventors across them, rather than diving into the document content where ideas are discussed~\cite{narin1985technology,ahmadpoor2017dual}. Documents entail many ideas, and linkages across them loosely capture which intellectual innovation is in focus and being transferred.  

By contrast, we conceptualize knowledge transfer in terms of scientific concepts, rather than documents associated to particular desirable outcomes, and demonstrate the importance of our derived features in knowledge transfer through machine learning model in a large-scale original dataset.

\begin{figure}[t]
\vspace{-0.3cm}
  \centering
      \begin{subfigure}[New concepts each year]{
        \label{fig:overtimeburnin}
        \includegraphics[width=0.47\linewidth]{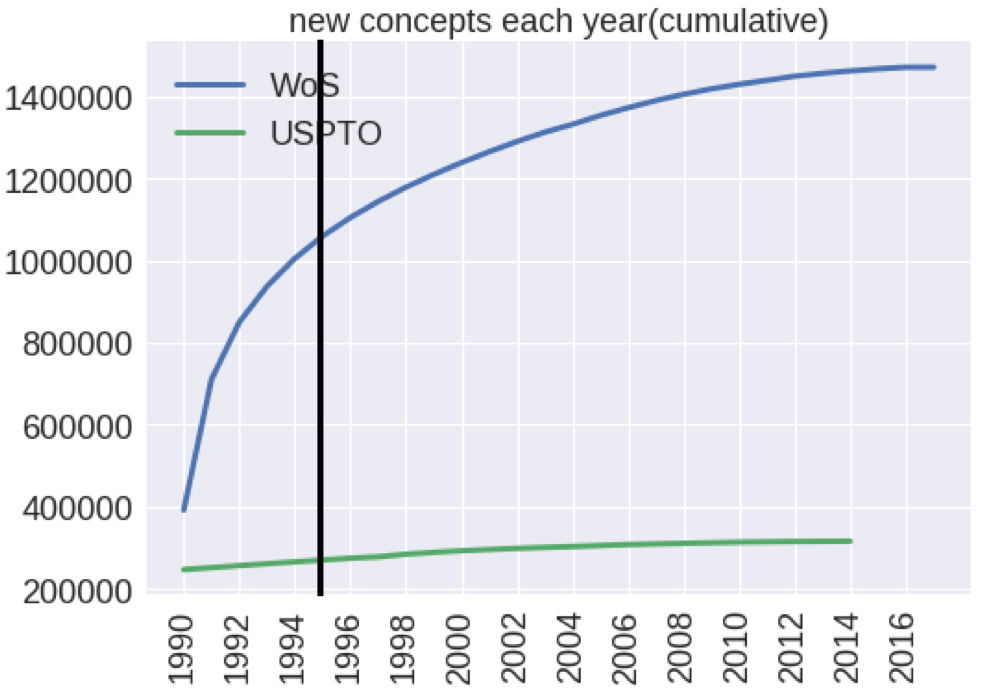}}
       \end{subfigure}
       \begin{subfigure} [\# transferred concepts]{
        \label{fig:overtimeyear}
        \includegraphics[width=0.43\linewidth]{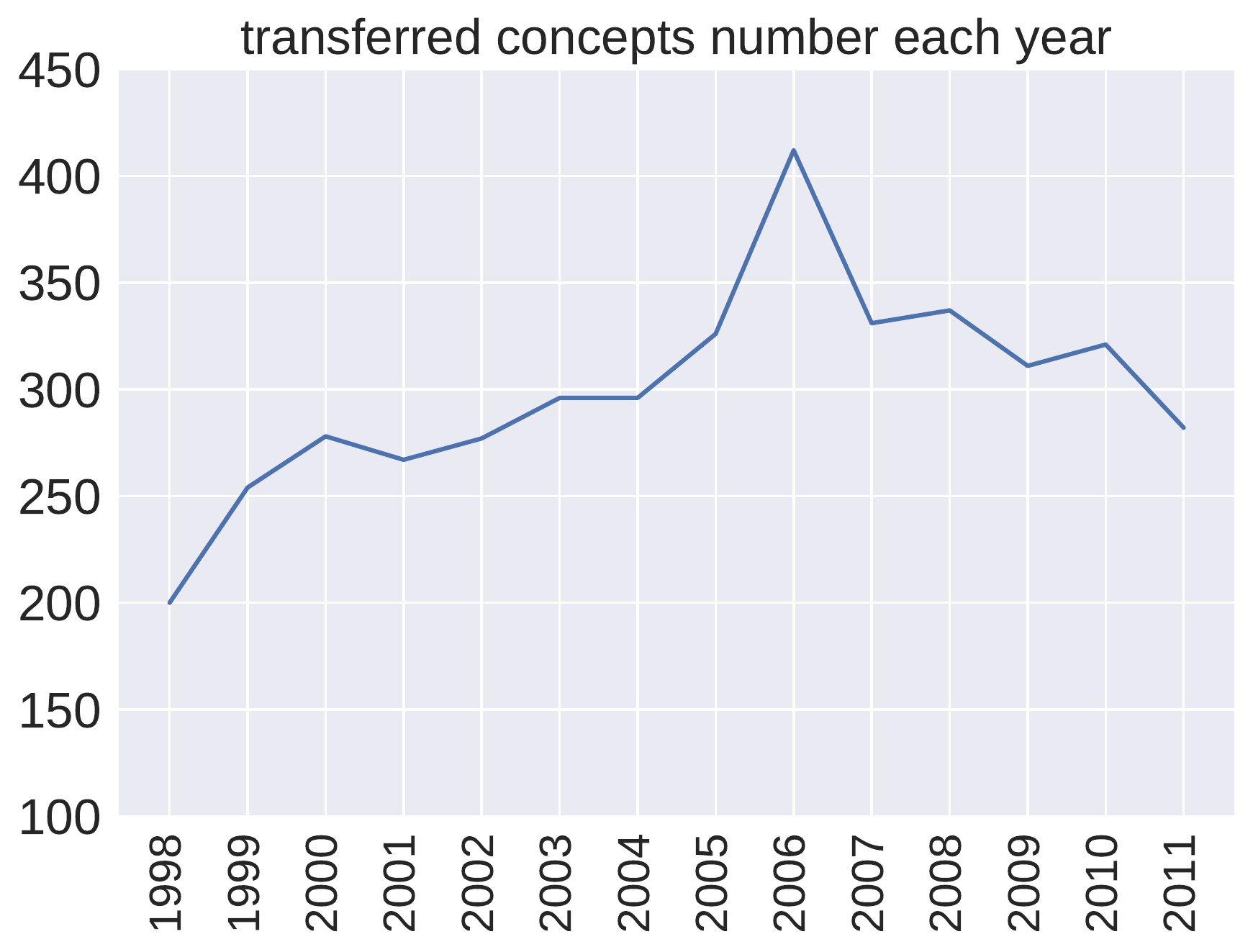}}
       \end{subfigure}
  \vspace*{-3mm}
  \caption{\textbf{Concepts' knowledge transfer over time}}
  \label{fig:overtime}
  \vspace*{-0.4cm}
\end{figure}

In this paper, we focus on studying patterns behind knowledge transfer from academia to research. We use ``knowledge transfer from academia to research'' in our study to mean a ``concept's transfer from research papers to patent documents/clinical trials'', or a concept that first appear in academia later get used a non-trivial frequency (decided by a pre-defined threshold) in practical outlets (patents, clinical trials). \footnote{We use knowledge transfer, concept transfer and idea transfer interchangeably throughout the paper}
In scientific writing, a scientific concept is a term or set of terms that have semantically coherent usage and reflect scientific entities – e.g., curricula, tools, programs, ideas, theories, substances, methods, processes, and propositions, which are argued to be the basic units of scientific discovery and advance \cite{toulmin1human}. We use the titles and abstracts of 38.6 million academic publications from the Web of Science (WoS) to identify 0.45 million new scientific concepts emerging between 1995 to 2014 through state of the art phrase mining techniques (AutoPhrase), and follow their trajectories in 4 million patent documents of the United States Patent and Trademark Office (USPTO), and 0.28 million clinical trials from U.S. National Library of Medicine.

In our analysis, we compare the properties of new scientific concepts that successfully transfer into patents with those that did not. We find that (a) the intrinsic properties of ideas and their temporal behavior, and (b) relative position of the ideas are the two mechanisms that determine whether an idea could transfer successfully. % need to incorporate intrinsic properties
In particular, we find new engineering-focused scientific concepts situated in emotionally positive contexts are more likely to transfer than other concepts. Furthermore, increased scientific hype and adoption across scientists, as well as usage in interdisciplinary venues over time, are early signs of impending knowledge transfer into technological inventions. Finally, we find that new concepts positioned close to concepts that already transferred into patents are far more likely to transfer than their counterparts. Based on the derived features, we further built model to predict the likelihood of knowledge transfer from papers to patents/clinical trials at individual concept level, and demonstrated our derived feature can achieve great performance, indicating that our proposed features can explain majority of the knowledge transfer cases.

\noindent
\textbf{Contributions  } Our main contributions are summarized as follows: 
(1) To the best of our knowledge, we present the first ever research that aims at understanding knowledge transfer at a large scale, using multiple corpora. 
(2) We are the first to leverage text mining techniques to understand transfer on scientific concept level, rather than document level. 
(3) We systematically analyzed the differences between transferable and non-transferable concepts, and identified the key mechanisms behind knowledge transfer. We showed our derived insights can help explain and predict knowledge transfer with high accuracy.

\begin{table}[t]
\vspace{-0.1cm}
\caption{\textbf{Examples of extracted scientific concepts}}
 \vspace*{-1mm}
 \scalebox{0.68}{
\begin{tabular}{cc}
\toprule
\textbf{Transferred Concepts} & \textbf{Non-transferred Concepts}                    \\ 
\midrule
Internet, world wide web,  & ethnographic exploration, \\ interactive visualization, web server, & immersive virtual reality, \\ 
gpu, recombinant protein production, &  european maize,  \\  
hcci engine, cloud service, & institutional demand,  \\
artificial magnetic conductor&automatic imitation,\\
multifunctional enzym, & network reorganization, \\
tissue remodeling, & human capital,\\
single photon detector & amercian theatre \\
\bottomrule 
\end{tabular}}

\vspace{-0.3cm}
\label{table:concept_example}
\end{table}

\section{Data Preparation and Processing}

In this section we introduce the dataset used in our study (Sec.~\ref{ssec:data_collection}), and present the concept extraction process (Sec.~\ref{ssec:concept_extraction}).

\vspace{-0.2cm}
\subsection{Collection of Text Corpora}
\label{ssec:data_collection}

\noindent
\textbf{Research Papers from WoS.}
We used scientific concepts extracted from Web of Science (WoS) as representation of knowledge in academia. We use the complete corpus from WoS (1900-2017) totaling 38,578,016 papers. %and most are in 1985-2017. 

% \subsection{Patents from United States Patent and Trademark Office}
\para{Patent Documents from USPTO.}
We used concepts extracted from 4,721,590 patents in the United States Patent and Trademark Office (USPTO) from 1976 to 2014 to represent general knowledge in the application domain.

\para{Clinical Trials.}
We used concepts extracted from 279,195 clinical trials from U.S. National Library of Medicine\footnote{Retrieved from clinicaltrials.gov} (1900 to 2018) to represent bio \& health sciences knowledge used in practice.

More details of the leveraged datasets are further elaborated in Appendix A. Note that our study inevitably suffer from data bias. For instance, not all practitioners will patent their idea, or file clinical trials, and that some clinical trails and patents are unused, thus there will be some false positives and negatives of `transferred' labels through our approach. Yet so far patent and clinical trial have been demonstrated to be the best proxy to study translational science from research to practice \cite{ahmadpoor2017dual}. Moreover, we have tried our best to mitigate such bias by investigating transfer patterns in both patent-heavy and patent-light fields, where we found very similar patterns emerge.

\subsection{Scientific Concept Extraction}
\label{ssec:concept_extraction}
Using titles and abstracts of articles, patents and clinical trials, we employ phrase detection technique AutoPhrase~\cite{shang2018automated}, to identify key concepts in the two corpora and trace their emergence and transfer across domains over time. Phrase detection identifies 1,471,168 concepts for research papers, 316,442 concepts for patents, and 112,389 concepts for clinical trials. Some samples of transferred concepts and non-transferred concepts extracted from WoS and USPTO by phrase detection are shown in Table. \ref{table:concept_example}. We observe that phrase detection results in high-quality concepts (92\% are labelled as high quality through our evaluations) that are suitable to investigate knowledge transfer across domains. Details of the phrase detection techniques, cleaning and evaluations are further discussed in Appendix B.

\para{New Concept Identification.}
The focus of this study is on new concepts and their careers. However, our sample of 1.5 million distinct concepts occur at any time in the corpus, some of which emerged long ago and others more recent. To avoid left-censoring issue (certain concepts appear before the start time of the recorded data thus we do not fully observe their behaviors) and identify `real' new concepts, we aggregate (or ``burn in'') the set of concepts over time, and count the number of new concepts that arrive each year. Early papers (starting 1900) identify many new concepts, but this quickly decelerates by around 1995 and then assumes a linear growth in vocabulary afterwards (see Fig. \ref{fig:overtimeburnin}). To identify that point, we aggregate the set of concepts every year with prior years until the rate of new concepts' introduction is approximately linear and stable. The point occurs after 1995, when 0.45 million scientific concepts are left. Then we follow knowledge transfer via these new scientific concepts, and find only $\sim$2.7\% of all concepts get transferred to patent, and only $\sim$11.3\% of bio \& health concepts get transferred to clinical trials across years. The number of transferred concepts each year from WoS to USPTO is illustrated in Fig.~\ref{fig:overtimeyear}.

\section{Feature Creation and Analysis}
%\xiang{try to cut down the writing by 30\%, and move minor details to appendix.}

Based on the concepts extracted from research papers, patent and clinical trial documents, we first create concept level features as motivated by prior literature on knowledge diffusion, and present a large-scale data analysis on transferred and non-transferred concepts to better understand properties facilitating the knowledge transfer process. Here we present transfer patterns from research paper to patent and omit clinical trial due to page limit \footnote{We find very similar transfer patterns emerge from research paper to clinical trial.}. 

\subsection{Intrinsic Properties of Concepts}
Motivated by previous works on knowledge diffusion and transfer, we extracted intrinsic concept features that would most likely facilitate a scientific idea's transfer into technological inventions, which can be classified into four categories: 1) hype features \cite{latour1987science,rossiter1993matthew}, 2) bridge positioning features \cite{shi2010citing,kim2017modeling}, 3) ideational conditions \cite{berger2005idea}, and 4) technological resonance \cite{narin1985technology}\footnote{While these features are not exhaustive, to the best of our knowledge they are the key factors most salient to the knowledge diffusion as discussed in literature}. The four sets of features represent the characteristic of individual concept from diverse angles, and as we will show signify the differences between transferred and non-transferred concepts, both in mean value (Appendix E) and temporal behavior.

\begin{figure}[H]
\vspace{-0.2cm}
  \centering
      \begin{subfigure}[Adopter size]{
        \label{fig:home_time}
        \includegraphics[width=0.47\linewidth]{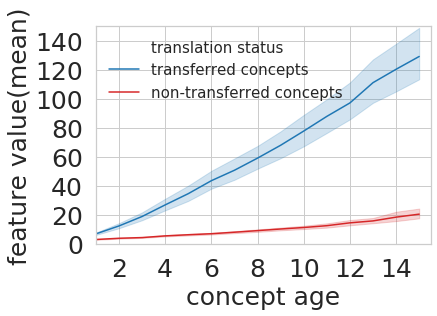}}
       \end{subfigure}
       \begin{subfigure} [Author repeated usage]{
        \label{fig:work_time}
        \includegraphics[width=0.47\linewidth]{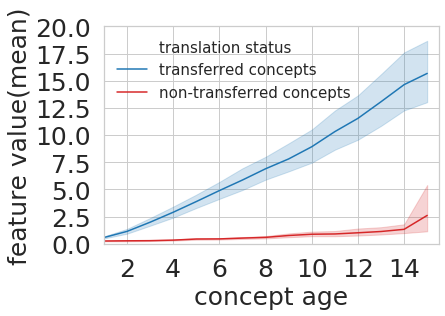}}
       \end{subfigure}
  \vspace*{-5mm}
  \caption{\textbf{Hype features.}}
  \label{fig:team}
  \vspace*{-4mm}
\end{figure}

\begin{figure}[hb]
\vspace{-0.3cm}
  \centering
       \begin{subfigure} [Discipline diversity]{
        \label{fig:work_time}
        \includegraphics[width=0.47\linewidth]{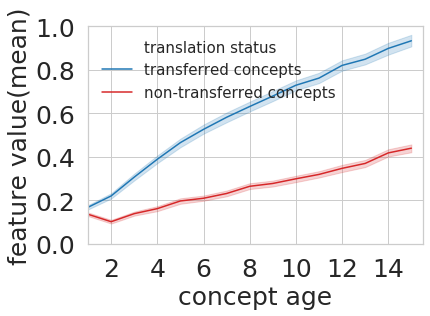}}
       \end{subfigure}
       \begin{subfigure} [Engineering Focus]{
        \label{fig:work_time}
        \includegraphics[width=0.47\linewidth]{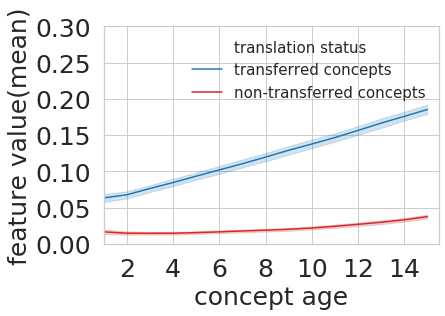}}
       \end{subfigure}
  \vspace*{-4mm}
  \caption{\textbf{Bridge positioning features.}}
  \label{fig:interdisciplinary}
  \vspace{-0.4cm}
\end{figure}

\begin{figure}[H]
\vspace{-0.3cm}
  \centering
      \begin{subfigure}[Emotionality]{
        \label{fig:home_time}
        \includegraphics[width=0.47\linewidth]{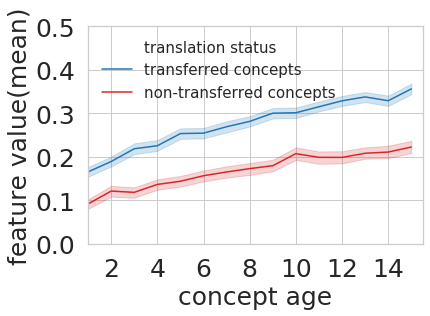}}
       \end{subfigure}
       %\begin{subfigure} [Emotionality]{
        %\label{fig:work_time}
        %\includegraphics[width=0.31\linewidth]{figs/emotion_prec1.png}}
       %\end{subfigure}
       \begin{subfigure} [Accessibility]{
        \label{fig:work_time}
        \includegraphics[width=0.47\linewidth]{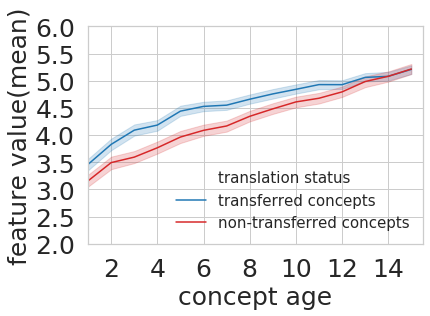}}
       \end{subfigure}
  \vspace*{-4mm}
  \caption{\textbf{Ideational conditions features.}}
  \label{fig:ideational}
  \vspace*{-4mm}
\end{figure}

\begin{figure}[t]
\vspace{-0.3cm}
  \centering
      \begin{subfigure}[Journal Linkage]{
        \label{fig:home_time}
        \includegraphics[width=0.47\linewidth]{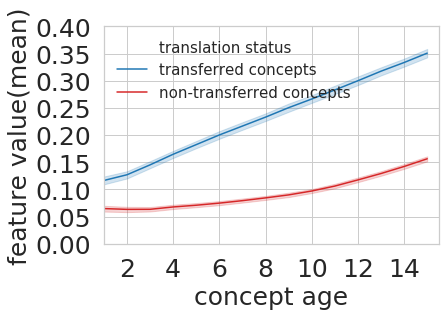}}
       \end{subfigure}
       \begin{subfigure} [University-Industry Relationship]{
        \label{fig:work_time}
        \includegraphics[width=0.47\linewidth]{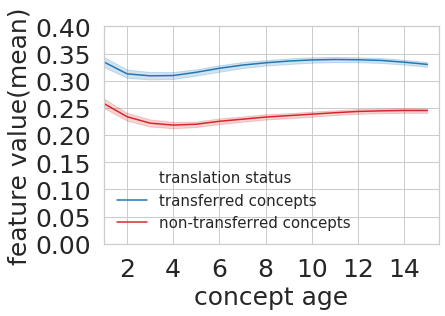}}
       \end{subfigure}
  \vspace*{-4mm}
  \caption{\textbf{Science technology linkage.}}
  \label{fig:transfer}
  \vspace*{-4mm}
\end{figure}

To illustrate concepts' temporal behavior over time, we plot the feature curves of transferred and non-transferred concepts over concept age. Details could be found in Figs. \ref{fig:team}-\ref{fig:position}.

\noindent
\textbf{Hype.}
This group of features draws on prior work concerning concept hype \cite{latour1987science, acharya2014rise, lariviere2014elite}. We include two features: the adopter size using the concept, and the degree to which authors repeatedly use the concept. We measure \textit{adopter size} as the total number of authors who employ a concept in a particular year, and \textit{author repeated usage} as the total number of previous authors continuing to use the concept. 

We found that transferred concepts generally demonstrate higher numbers of adopters and repeated usage. Furthermore, transferred concepts attract adopters at a faster rate than non-transferred concepts. We also found that transferred concepts are repeatedly used much more often by the previous authors when controlling for concept age. What's more, we observe an increasing gap with regard to `hype' features between transferred and non-transferred concepts over time, possibly due to the preferential attachment effect \cite{newman2001clustering}.

\para{Bridge Positioning.}
This group of features identify the disciplinary placement of concepts. 
Previous works argue that knowledge transfer is facilitated when ideas are placed at the boundary of fields and in fields especially relevant to technological invention (\cite{shi2010citing}). Here we include two features: \textit{discipline diversity} and \textit{engineering relation} in this group. \textit{Discipline diversity} is computed as a concept’s average entropy across NRC discipline subject codes (sociology, math, economics, etc.), and \textit{engineering relation} is computed as the proportion of engineering fields among all the fields using the concept. 

We found transferred concepts are more likely to be used in interdisciplinary and engineering venues. Moreover, transferred concepts gained greater interdisciplinary attention over time compared to non-transferred concepts, as shown in Fig. \ref{fig:interdisciplinary}. The finding is consistent with the assumption that transferred concepts are likely to achieve a more diverse audience than non-transferred concepts. Engineering focused concepts also achieved a higher knowledge transfer rate, which supports our hypothesis that knowledge transfer is facilitated when ideas are placed at the boundary of fields especially relevant to technological invention like engineering (e.g. mechanical engineering). Once again, we observed the difference of `bridge positioning' feature values between transferred and non-transferred concepts increase over time.

\para{Ideational Conditions.}
This group of features represents the semantic context and expression of a concept. How the concept is related to other concepts and the style with which the concept is expressed can both influence the diffusion and transfer process \cite{hamilton2016diachronic}. Here we select \textit{emotionality}, and \textit{accessibility} in this group, and calculated them through LIWC and Dale Chall metric (details in Appendix C).

We found transferred concepts are embedded in more emotional context, and described in more difficult language, compared to non-transferred counterparts. In a similar way, we plot ideational condition features over time for transferred concepts and non-transferred concepts in Fig. \ref{fig:ideational}. We found that transferred concepts were consistently placed in increasingly positive contexts and conveyed in more difficult language over time, compared to non-transferred concepts, although the accessibility gap decreases over time. 

\para{Technological Resonance.}
This group of features quantifies the extent to which a concept is established within an environment conducive to link scientific publications with patents and other outcomes 
\cite{narin1985technology,tijssen2001global}. We measure this as \textit{journal linkage} and \textit{university-industry relationship} in our study. \textit{journal linkage} is computed as the percentage of journals where the concept is situated that have been cited by patents before. \textit{university-industry relationship} is calculated as the proportion of industry-affiliated authors out of all the authors employing the term each year. Should a scientific concept be in a high bridging space like these, they will more likely transfer.

Transferred concepts are more likely to be mentioned in journals that have been cited by patents, and this relationship strengthens over time. We also find that if a concept is associated with more industry-affiliated authors, the concept has a higher potential to transfer. While the industry-affiliate author percentage between transferred and non-transferred concepts remain relatively stable, the gap between them with regard to journal linkage gets greater over time.

\subsection{Relative Position in Concept Co-occurrence Graph}
In addition to the above features, we investigate the same data with a relational approach \cite{hofstra2019diversity}. Intuitively, how a concept get positioned/co-used with other concepts may be associated with knowledge transfer.

\begin{figure}[t]
\vspace{-0.1cm}
    \centering
    \includegraphics[width=0.95\linewidth]{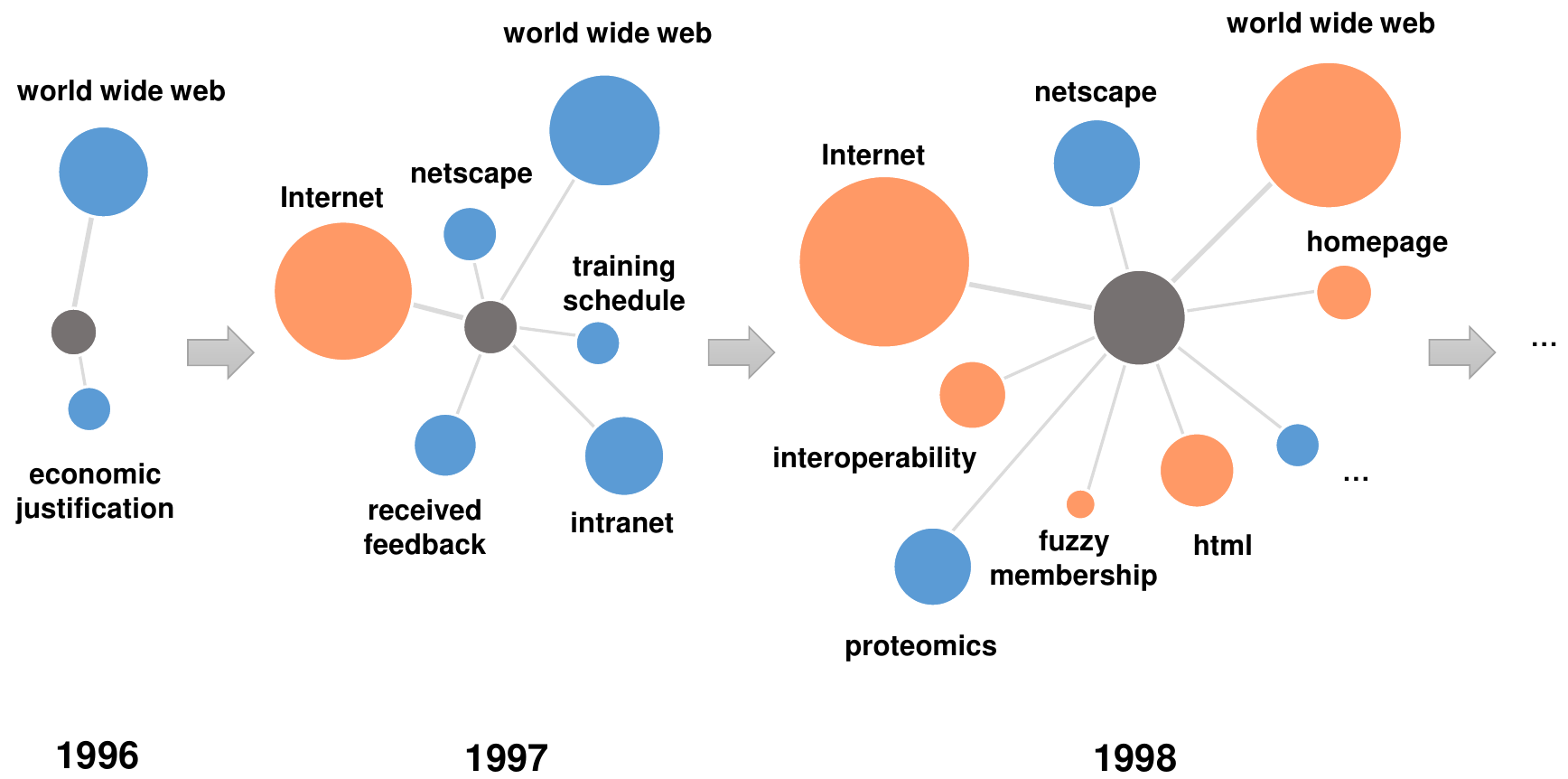}
    \vspace{-0.3cm}
    \caption{\textbf{Illustration of the dynamic graphs that capture interactions between \textit{search engine} and its co-occurrence concepts.} The orange circles denote transferred concepts while the blue denotes non-transferred ones; the circle size represents the node degree.}
    \label{fig:co-occurrence}
    \vspace{-0.3cm}
\end{figure}

As a motivating example, we plot the local co-occurrence network of concept \textit{search engine} in Fig. \ref{fig:co-occurrence}. The central grey circle is \textit{search engine}, and the orange nodes denote the transferred neighbor concepts while the blue denotes the non-transferred one. \textit{Search engine} first emerged in WoS in 1992 and entered USPTO in 1998. Coincidentally, the percentage of its transferred neighbors increased rapidly right before 1998, which indicates the neighboring concepts that get co-used with a concept may embed useful signals that explain concept transfer. The consistency between co-occurrence network and transfer status is also common in other concepts. 

\begin{figure}[t]
\vspace{-3mm}
  \centering
      \begin{subfigure}[Weighted Degree]{
        \label{fig:degree}
        \includegraphics[width=0.47\linewidth]{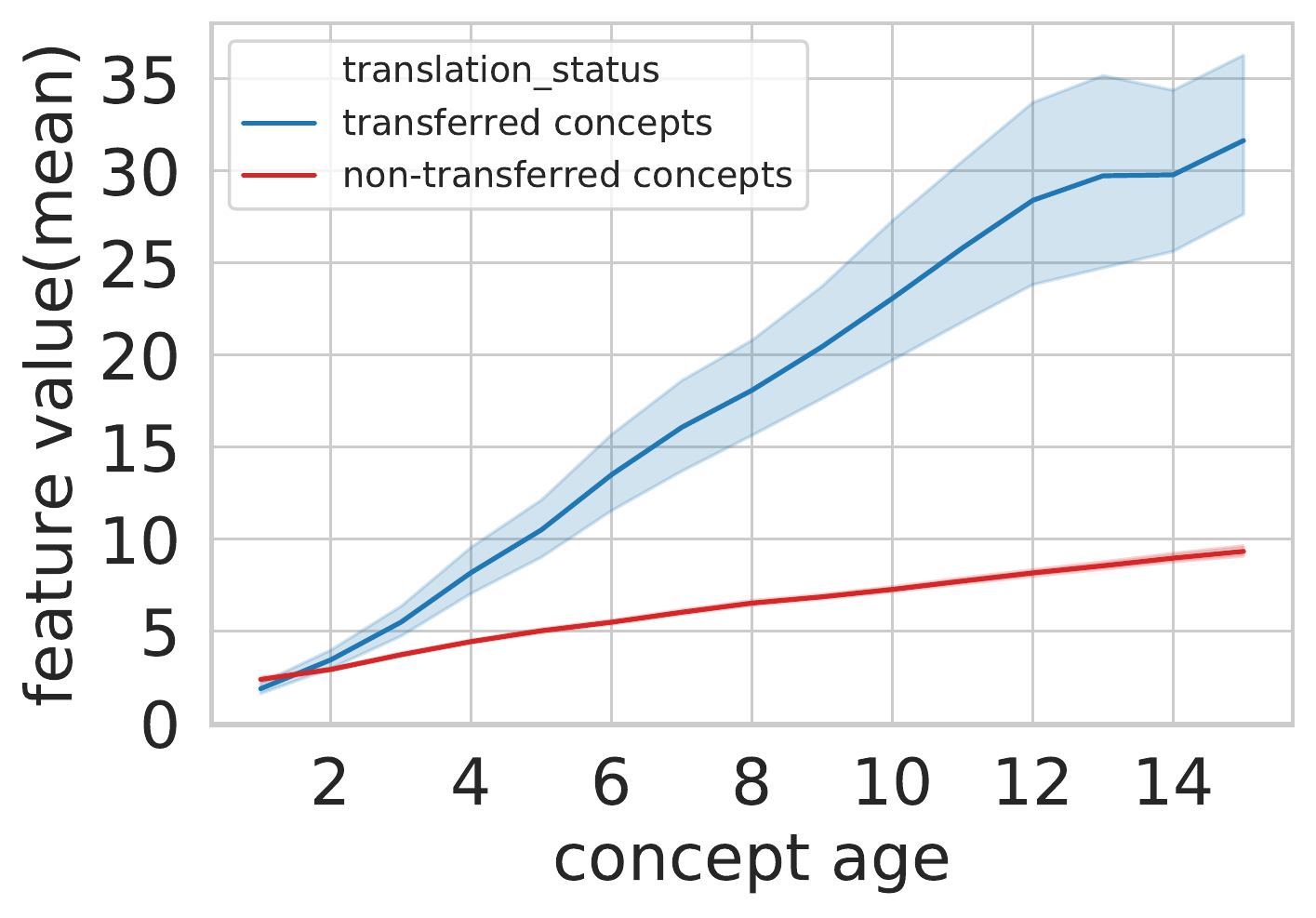}}
       \end{subfigure}
       \begin{subfigure} [Weighted Percentage]{
        \label{fig:pctg}
        \includegraphics[width=0.47\linewidth]{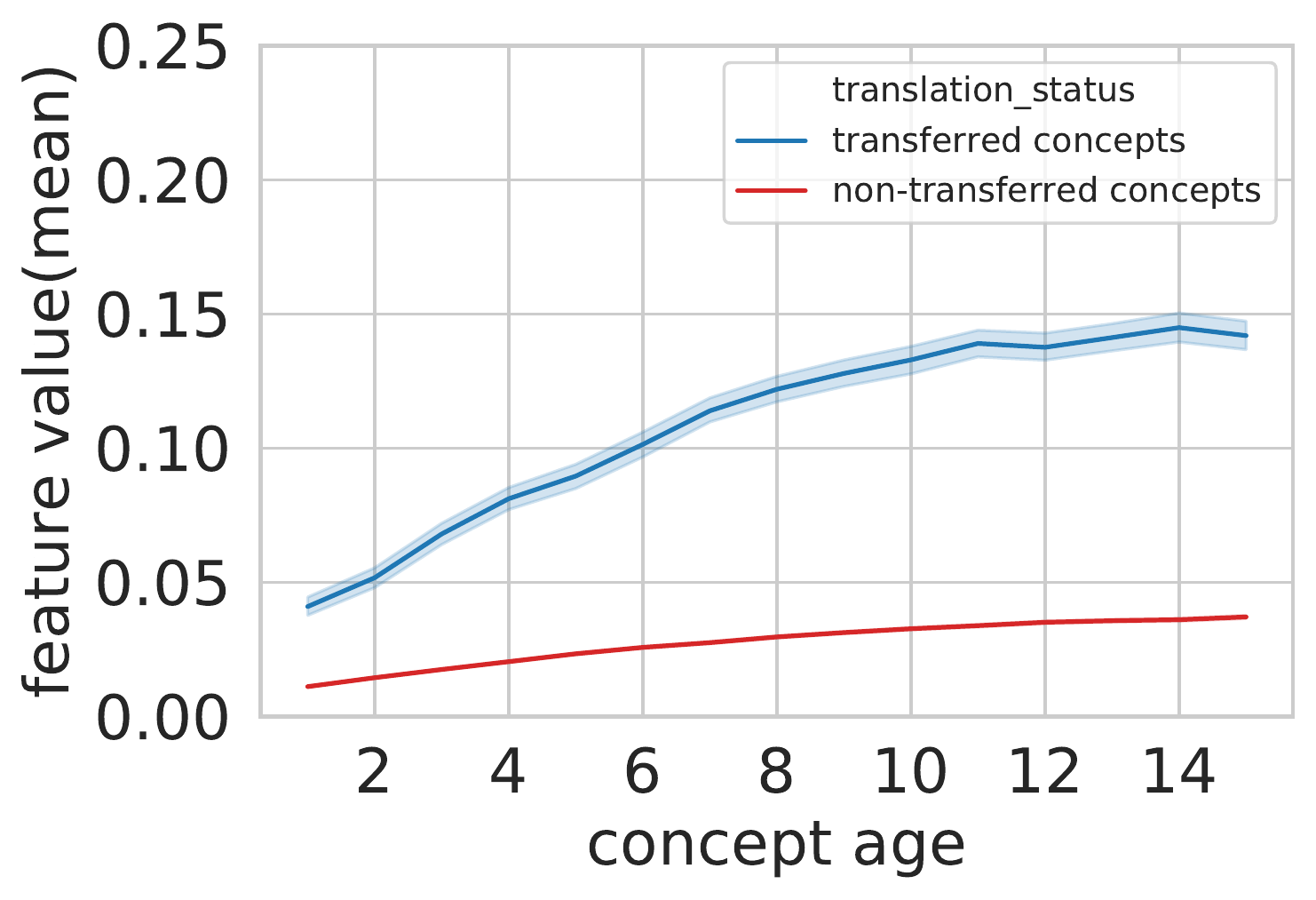}}
       \end{subfigure}
  \vspace*{-4mm}
  \caption{\textbf{Graph features.}}
  \label{fig:position}
  \vspace{-4mm}
\end{figure}

 To facilitate analysis, we construct a dynamic graph $\mathcal{G}$ for concept co-occurrence. Each node in graph denotes a concept which has occurred in the corpus. Each edge between two nodes indicates the two concepts co-occur in at least one document in the corpus, and we define the edge weight as the number of documents the two concepts co-occur. We sort all documents by year and construct a graph at each time-stamp, then we will get a set of graphs $\{\mathcal{G}\}=\{\mathcal{G}^{(1)},\cdots,\mathcal{G}^{(t)}\}$ as \textit{dynamic concept co-occurrence graph}. This set of graphs reflects the dynamic succession of concepts' neighbors and provides us with extra temporal information on local graph structures. Based on dynamic concept co-occurrence network, we derived two graph features: \textit{weighted degree} and \textit{weighted percentage of transferred neighbors} as specified in Appendix D.

The curves of the two features over time are shown in Fig. ~\ref{fig:degree} and Fig. ~\ref{fig:pctg}.
We find that transferred concepts indeeed have higher weighted degrees and weighted percentages compared to non-transferred ones, which indicates the importance of utilizing concept co-occurrence for knowledge transfer prediction.

\para{Field Comparison \& Feature Correlation } We further carried out analysis on feature correlation, and comparison across fields, which is discussed in detail in Appendix F and Appendix G.

\para{Summary.} Results of our data analysis support the conclusion that knowledge transfer is not by chance but follows specific patterns. Whether a concept will transfer from research to practice in the immediate future depends largely on their (a) individual properties over time, and (b) relative positions with respect to other concepts. 

\section{Predictive Analysis of Features}

So far we have systematically analyzed the potential factors that reflect the process of knowledge transfer from research to practice. But how well can these features explain and predict knowledge transfer in practice? In this section, we seek to shed light on this question through predictive analysis.

\subsection{Prediction Task Formulation}

Will a scientific concept transfer from academic papers to patent documents in the next X years? Here we consider the predictive task which aims to predict concept transfer status given all observed historical data. As there can be only two potential outcomes --- either the concept transfers or not --- the proposed prediction task is essentially a binary classification problem. We label a concept as transferred if it first originates in research papers and later get used at least 5 times in practical outlets (patents, clinical trials) within X years after the concept's birth in research papers.

\begin{figure}[t]
    \centering
    \includegraphics[width=1.0\linewidth]{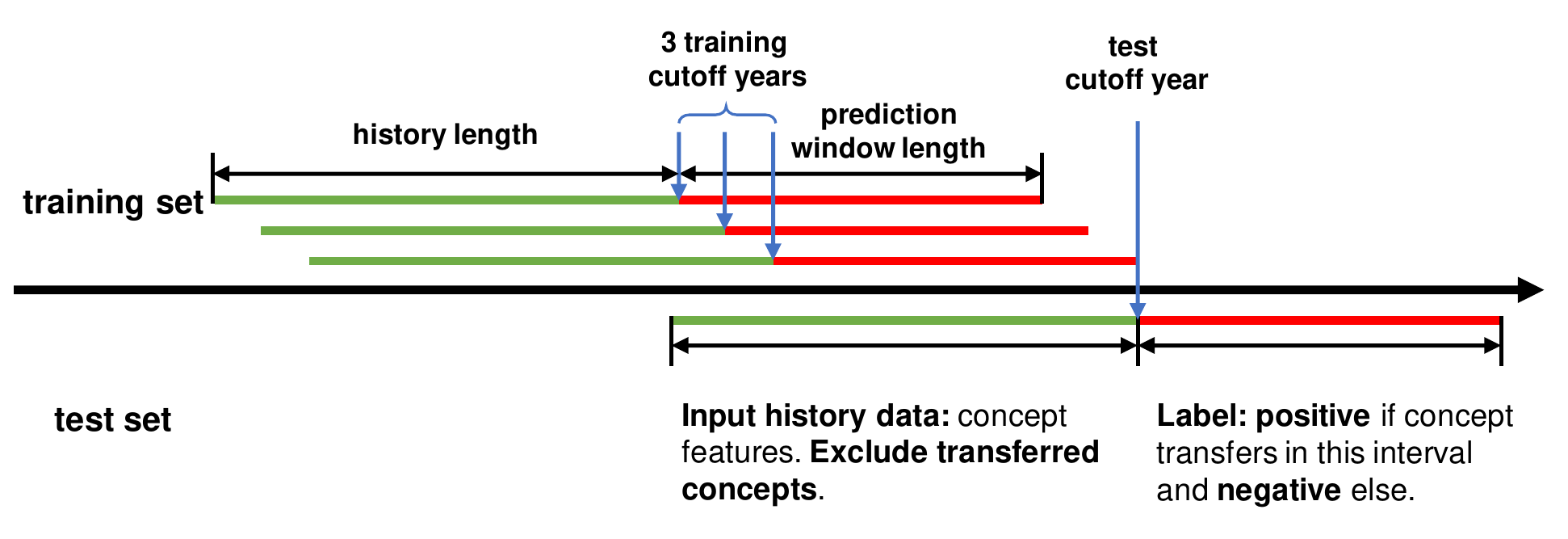}
    \vspace{-0.6cm}
    \caption{\textbf{Illustration of the concept's knowledge transfer prediction task and the partition of training/test sets.} The green lines denote historical input while the red lines denote prediction window. The prediction intervals of training and test sets should not overlap.}
    \label{fig:split}
    \vspace{-0.3cm}
\end{figure}

We denote all $N$ concepts' time-series attributes at one particular time-stamp as $X \in \mathbb{R}^{N\times N_x}$, where $N_x$ is the dimension of attributes.

As shown in Fig. ~\ref{fig:split}, the goal of the transfer prediction problem is to construct a function $f(\cdot)$ mapping historical time-series attributes to the future transfer probability of concept,
%$$f:\left[ \left(\textbf{x}^{(t-k)}_i,\mathcal{G}^{(t-k)}\right),\cdots,\left(\textbf{x}^{(t-1)}_i,\mathcal{G}^{(t-1)}\right) \right] \rightarrow P\left(y_i^{(t)}=1\middle|\;\cdot\right) $$
$$f:\left[ \left(\textbf{x}^{(t-k)}_i\right),\cdots,\left(\textbf{x}^{(t-1)}_i\right) \right] \rightarrow P\left(y_i^{(t)}=1\middle|\;\cdot\right) $$
where $\textbf{x}_i=X_{i,:}$ denotes the attribute vector of concept $i$, $P\left(y_i^{(t)}\middle|\;\cdot\right)$ is the conditional probability and $k$ is input history length. $y_i$ denotes transfer status of concept $i$ in next $T$ years, \ie, the ground truth label of $y_i^{(t)}$ is 1 if it transfers in $t\sim t+T-1$ else 0, and $T$ denotes \textit{prediction window length}. Particularly, we note $t$ as \textit{cutoff year} and our model inputs the attributes previous to this time-stamp and predicts future transfer probability. For simplicity, we denote $P\left(y_i^{(t)}=1\middle|\;\cdot\right)$ as $p_i^{(t)}$.

Accordingly, if the true transfer status is $\mathrm{y}_i^{(t)}$, the loss function for cutoff year $t$ is 
%\begin{equation}
    $$\mathcal{L}=-\sum_{i} \left[\mathrm{y}_i^{(t)}\log p_i^{(t)}+ \left(1-\mathrm{y}_i^{(t)}\right)\log\left(1-p_i^{(t)}\right)\right]$$
%\end{equation}

\subsection{Prediction Models}

\para{Feature based Model.}
We use logistic regression (LR) as an interpretable model. To better validate our finding, we also run a mixed effects logistic regression detailed in Appendix I, a form of Generalized Linear Mixed Model (GLMM), to help explain variance both within-concept and across-concept. The results from the mixed effects logistic regression are nearly identical with our findings from the vanilla logistic regression, except for slight changes in the magnitude of coefficients, so we only report performance of LR in our analysis.

\para{Deep Sequence Model.}
To model a concept's temporal features, \ie time-series attributes, we further propose RNN sequence models. According to Sec.3, some time-series features are strongly related to potential transfer; therefore, we adopt Recurrent Neural Network (RNN) models (e.g., LSTM and GRU) which are built to capture temporal dependencies (Details in Appendix H).

\section{Experiments}
%\para{Experiment Setup.} 

\noindent
\textbf{Experiment Set-up.}
We apply Z-score normalization on time-series attributes and divide dataset into training/test sets as Fig.~\ref{fig:split} shows.
Given test cutoff year $t$, we first ensure the prediction intervals (red line in Fig. \ref{fig:split}) of training and test set have no overlap to avoid data leakage, and then use the latest three cutoff years as train cutoff years. For instance, if test cutoff year is 2008 and prediction window is 5 years long, the latest training prediction interval should be 2003$\sim$2007 and thus we use 2001, 2002, 2003 as training cutoff years. As concept transfer status is irreversible, we exclude all transferred concepts from test set but still use them to train. 

Details of model training and hyperparameter settings are discussed in Appendix J. Here we primarily report experimental results on knowledge transfer prediction from WoS to USPTO, while using clinical trial as a robustness check.

\para{Evaluation Metric.}

We adopt area-under-curve (AUC) as evaluation metric, which is not affected by data imbalance in test set.

\begin{table}[t]
\vspace{-0.2cm}
\caption{\textbf{Performances (mean AUC) of next-3-year transfer prediction for cutoff year 2008 from Web of Science to patent, and from Web of Science (Biology \& Health Sciences papers) to clinical trial. All results are generated by 3-run experiments.}}
\vspace{-0.3cm}
\scalebox{0.7}{
\begin{tabular}{lcccccc}
\toprule
 & \multicolumn{2}{c}{\textbf{Patent}} & \multicolumn{2}{c}{\textbf{Clinical Trial}}  \\
% \hline
\textbf{Method}~/~\textbf{History length}          & \textbf{3 years}    & \textbf{5 years}   & \textbf{3 years}  & \textbf{5 years}      \\ 
\midrule
LR w. graph features  &0.792      &0.792      & 0.656     & 0.661     \\
LR w. all features    &0.794      &0.800      & 0.675     & 0.677    \\
RNN w. graph features & 0.793     & 0.797     & 0.706     & 0.726      \\
RNN w. all features   & 0.803     & 0.809     & 0.715     & 0.734      \\
\bottomrule
\end{tabular}
}
\label{table:results}
\vspace{-0.3cm}
\end{table}

\subsection{Results} 
We first compared the performances of all aforementioned models for cutoff year 2008 on datasets: from WoS to patent, and from WoS bio \& health science papers to clinical trials\footnote{Note that for knowledge transfer to clinical trial, we excluded bridge positioning features since we focused on bio \& health science only.}. For each cutoff year, we ran two sets of experiments with training history lengths of 3 and 5 years and repeated 3 times for each experiment. The performances (mean AUC) are summarized in Table. \ref{table:results}.

\para{Patent vs. Clinical Trial}
As a robustness check, we tested our model on both knowledge transfer from WoS to patent, and to clinical trial. We obtained consistent main attribute importance results based upon clinical trial data.

As can be observed from Table. \ref{table:results}, our derived features achieve good result, i.e. AUC 0.80, in predicting knowledge transfer, demonstrating knowledge transfer can be largely explained by our proposed mechanisms.

\para{Study of Feature Importance.}

\begin{figure}[t!]
\vspace{-0.0cm}
    \centering
    \includegraphics[width=0.6\linewidth, angle =270]{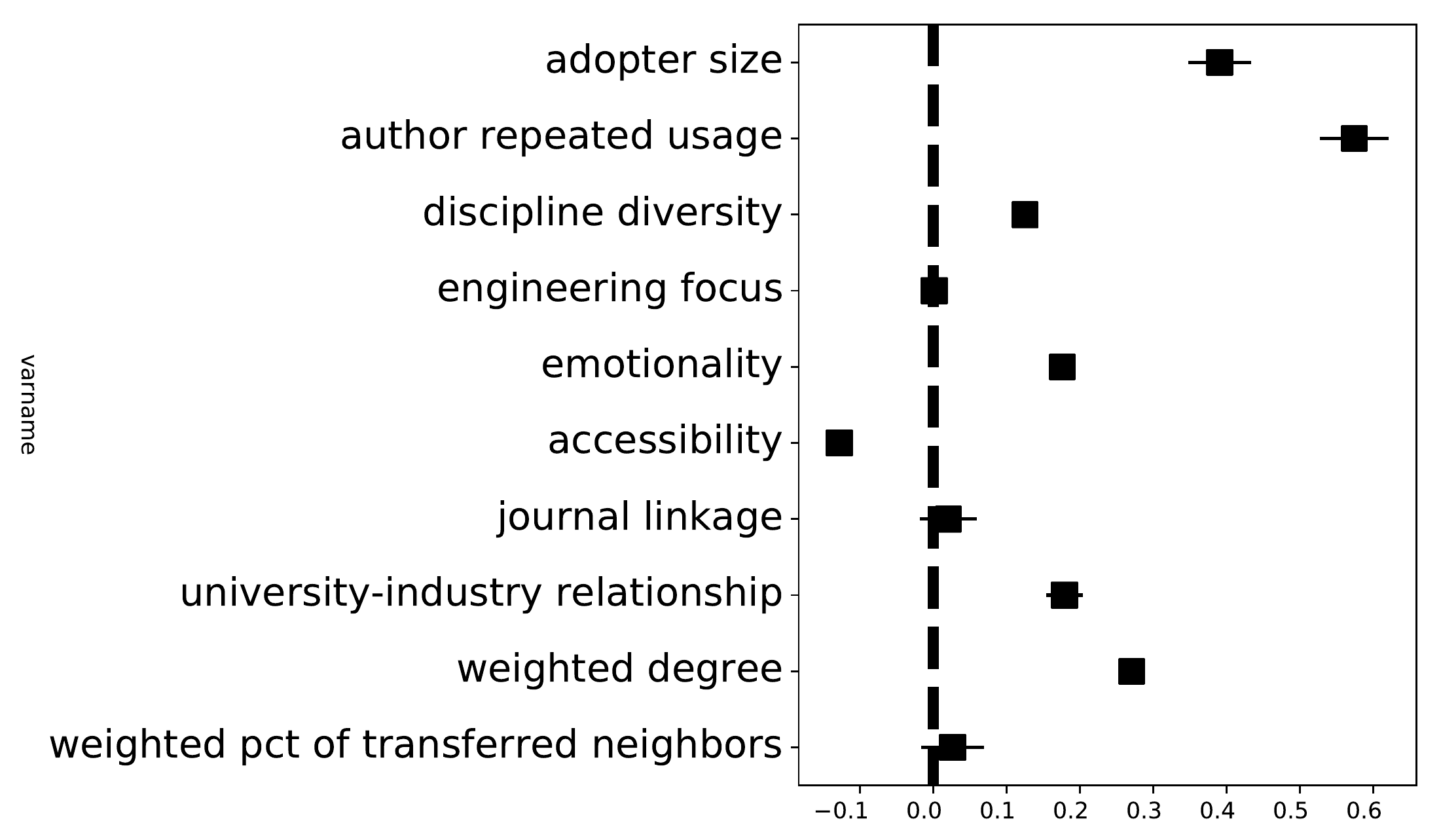}
    \vspace{-0.2cm}
    \caption{\small\textbf{Feature Importance Study} }
    \label{fig:featureimportance}
    \vspace{-0.4cm}
\end{figure}

In Fig. \ref{fig:featureimportance}, we further plot the standardized coefficients of each temporal feature from the logistic regression to understand how a specific attribute contributes to the knowledge transfer. We observed that author repeated usage, adopter size, and weighted graph degree are the three most important factors in influencing knowledge transfer. 

Next, we studied feature importance in our proposed models, where we ran both models with different sets of features on knowledge transfer from WoS to patent. The result is summarized in Table. \ref{table:features_results}. As reflected by experiment results with RNN model, graph features achieve best prediction results compared to other feature sets, followed by ``bridge positioning" features, ``ideational conditions'', and ``hype'' features,  suggesting that the relative position position of the concept in the semantics network is the single most useful feature set that explains concept transfer.

\para{Study on Field Difference}
We studied the prediction performance of the proposed model in different fields. We partitioned the concepts used in Web of Science based on their field, trained and tested models separately using 5-year historical data as training inputs with train cutoff year 2003 and tested cutoff year 2008 for next 3-year prediction. We observed that it is easiest to predict knowledge transfer from academia to practice in humanity (AUC 0.973), followed by physical \& math science (AUC 0.791), bio \& health science (AUC 0.783), engineering (AUC 0.782), social science (AUC 0.706) and agriculture (AUC 0.633), which indicates our proposed mechanism can explain knowledge transfer quite well in most fields other than agriculture.

\begin{table}[t]
\centering
\vspace{-0.1cm}
\caption{\textbf{Performance with different feature groups.}} 
\vspace{-0.2cm}
\scalebox{0.8}{
\begin{tabular}{@{\hspace{0.1cm}}lc@{\hspace{0.1cm}}}
\toprule
\textbf{Method}                               & \multicolumn{1}{l}{\textbf{AUC}} \\
\midrule
LR w. ``hype" features &  0.629   \\
LR w. ``bridge positioning" features &  0.681\\
LR w. ``ideational conditions" features  &  0.662 \\
LR w. ``sci-tech linkage" features &  0.670\\
LR w. graph features &  0.792 \\
LR w. all features   &  0.800 \\
RNN w. ``hype" features &  0.641    \\
RNN w. ``bridge positioning" features &  0.708\\
RNN w. ``ideational conditions" features  &  0.686 \\
RNN w. ``sci-tech linkage" features &  0.676\\
RNN w. graph features &  0.797 \\
RNN w. all features   &  0.809 \\

\bottomrule
\end{tabular}
}
\label{table:features_results}
\vspace{-0.0cm}
\end{table}

\subsection{Sensitivity Analysis}
Finally, we tested our proposed models under different settings on WoS to patent. We investigated whether our proposed transfer model is influenced as a result of 1) varying length of historical observations, 2) varying prediction time window, and 3) varying cutoff year.

\begin{figure}[t]
 \vspace{-0.1cm}
    \centering
    \includegraphics[width=0.9\linewidth]{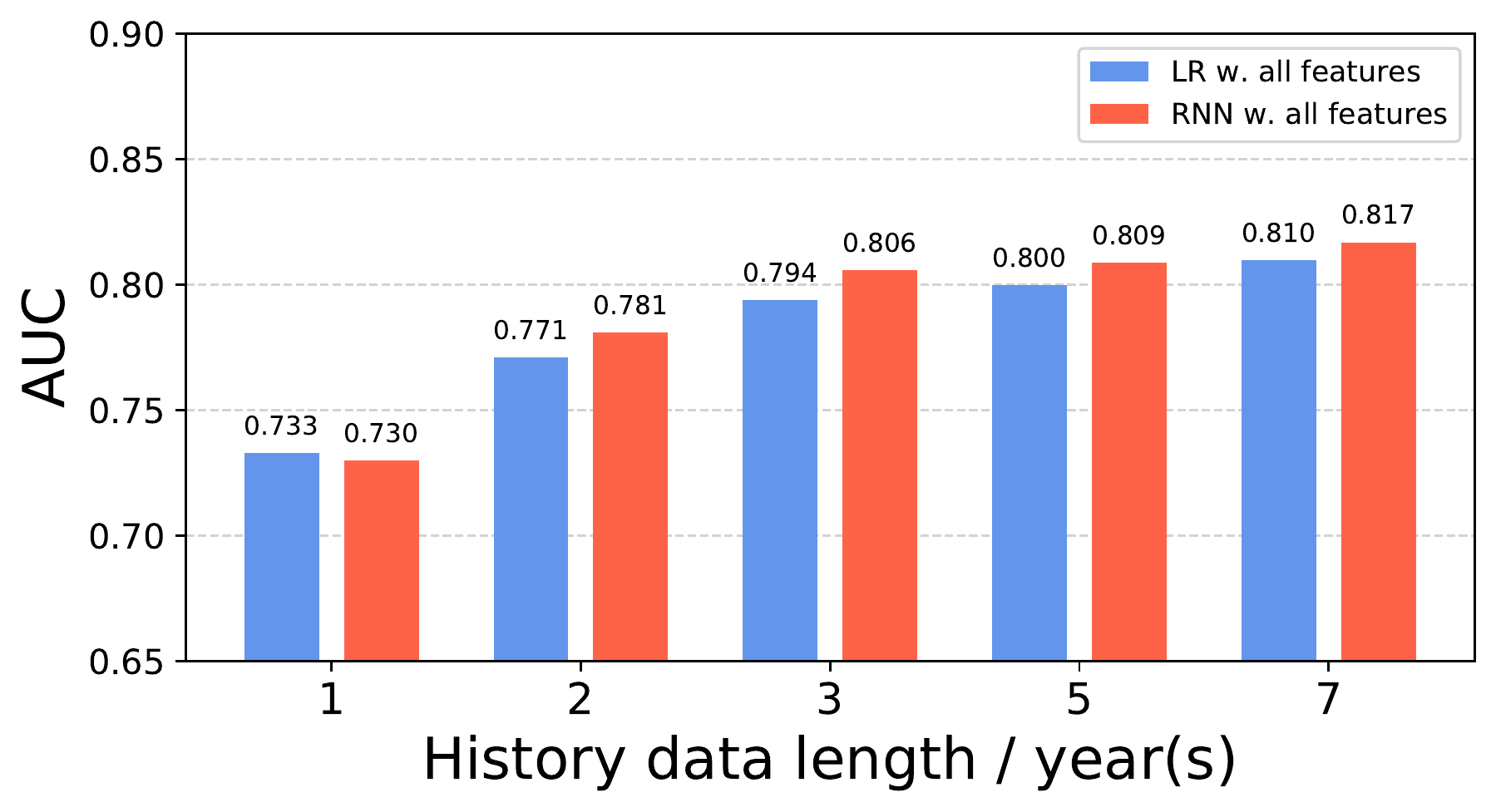}
    \vspace{-0.4cm}
    \caption{\textbf{Performances of next-3-year knowledge transfer prediction with cutoff year 2008.} We use different length of historical data as training data. }
    \label{fig:hist_len}
     \vspace{-0.4cm}
\end{figure}

\para{1. Length of Observation History.}
Fig. \ref{fig:hist_len} demonstrates the effects of historical observation length on performance, where we selected 1 year, 2 years, 3 years, 5 years and 7 years of observation before cutoff year 2008 as training sets. We found that the longer the observation data, the better prediction result we will get for the transfer prediction, which can be explained by the fact that longer observation better captures knowledge transfer patterns. We also note that performance starts to plateau when observation length gets larger, indicating that longer training sets only provide limited additional signal. All this indicates knowledge transfer is most influenced by behavior of concepts in the recent few years.

\begin{figure}[t!]
\vspace{-0.1cm}
    \centering
    \includegraphics[width=0.85\linewidth]{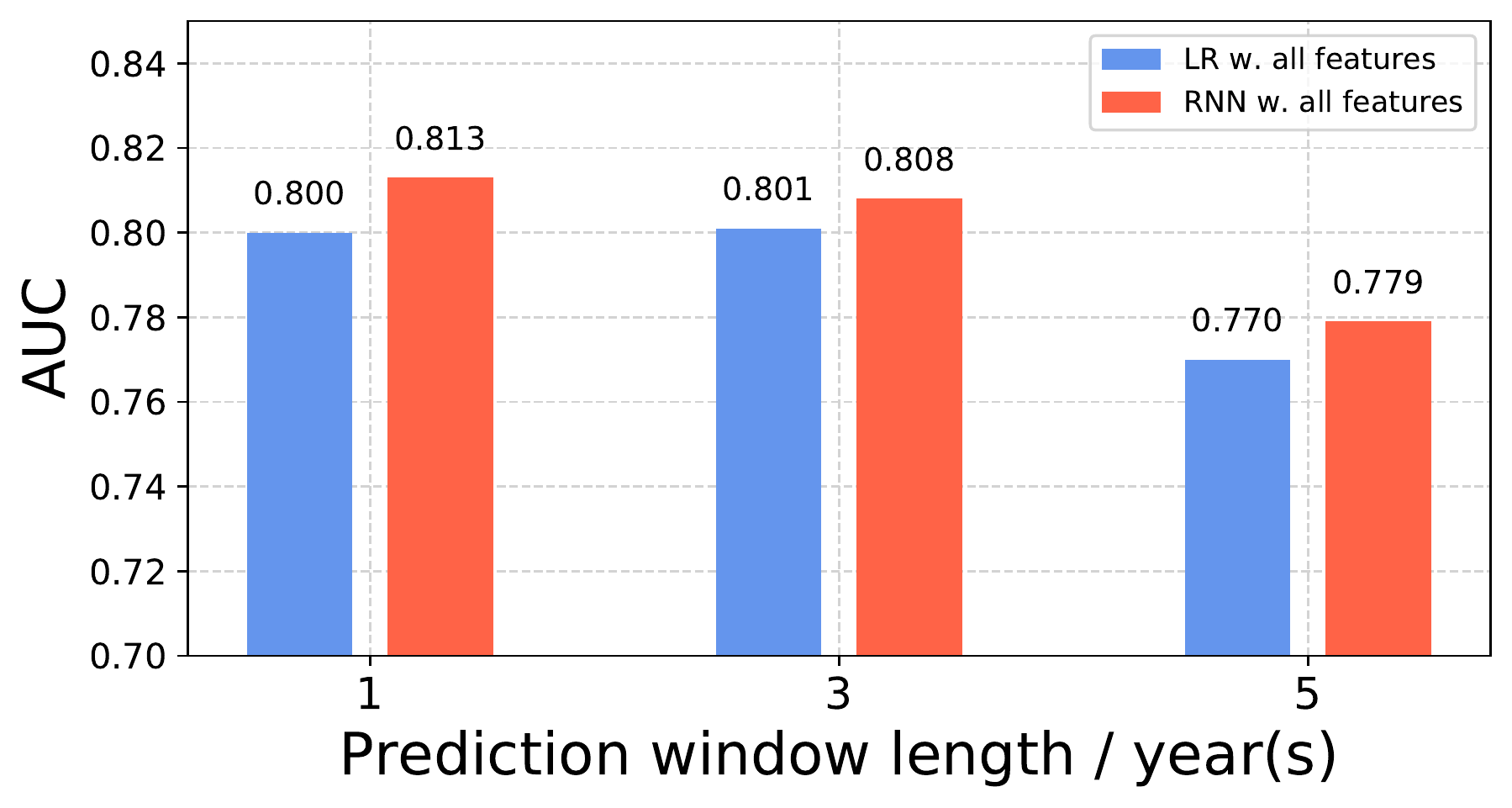}
    \vspace{-0.2cm}
    \caption{\small\textbf{Performances under different lengths of prediction time windows.} 5-year historical data was used for training with train cutoff year 2003 and test cutoff year 2008.}
    \label{fig:pred_len}
    \vspace{-0.4cm}
\end{figure}

\para{2. Length of Prediction Time Window.}
Fig. \ref{fig:pred_len} further illustrates the knowledge transfer prediction performance with prediction window of 1 year, 3 years and 5 years, representing the case when predicting whether a concept will transfer in next 1 year, 3 years or 5 years, respectively. To compare them fairly, we fix both training and testing cutoff years to keep time interval from training set to test set unchanged, which is different from the setting in previous experiments. As can be observed, prediction performance is consistently best when prediction window is 1 year, indicating the increasing difficulty in capturing long-term temporal pattern of knowledge transfer of our proposed mechanisms. 

\begin{figure}[t!]
\vspace{-0.1cm}
    \centering
    \includegraphics[width=0.85\linewidth]{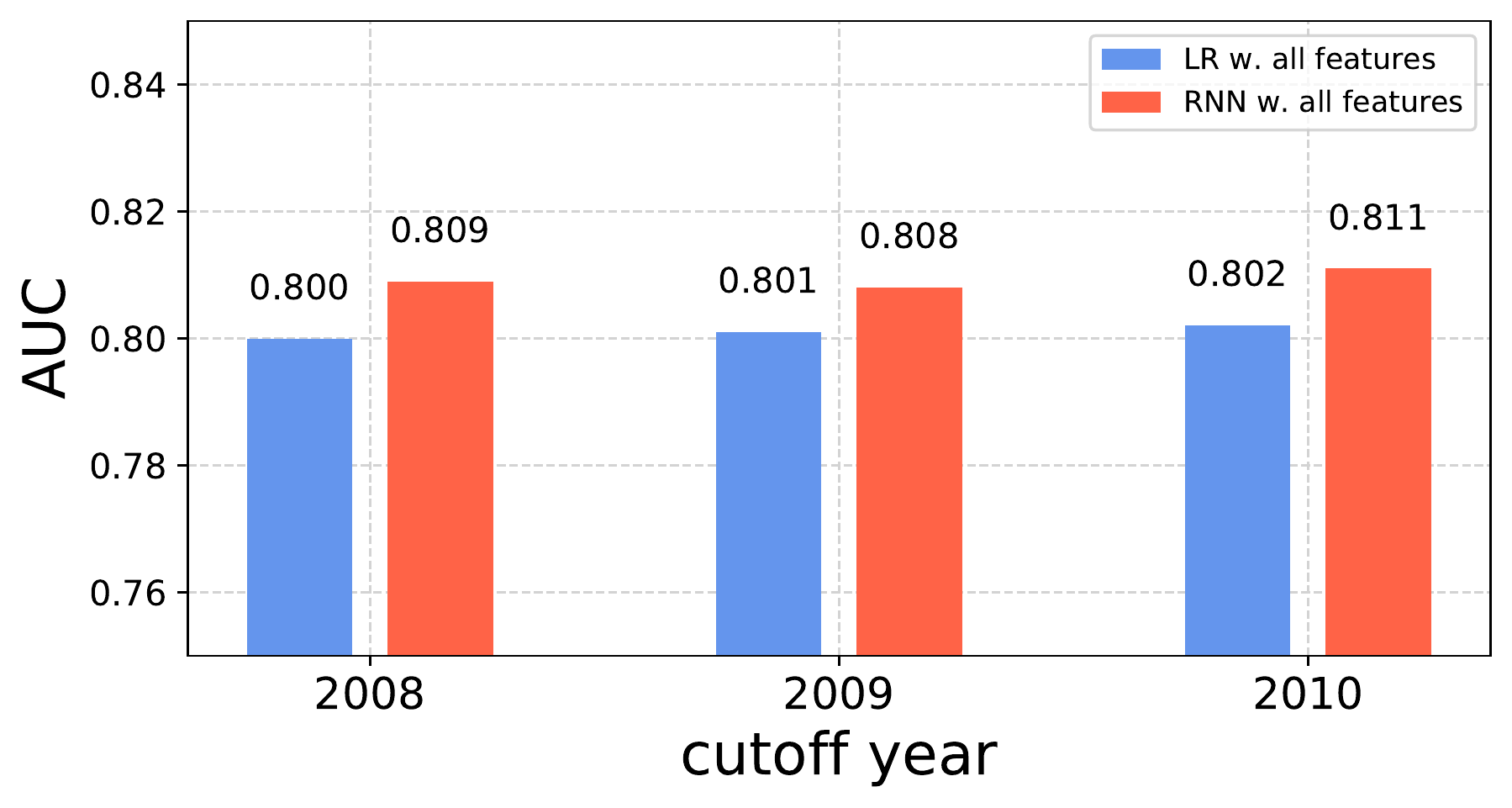}
    \vspace{-0.2cm}
    \caption{\small\textbf{Performances of next-3-year transfer prediction under different cutoff years.} We use 5-year historical data as training inputs.}
    \label{fig:cutoff}
    \vspace{-0.4cm}
\end{figure}

\para{3. Cutoff Year.}
We also tested our model with different cutoff years (i.e., 2008, 2009 and 2010), representing knowledge transfer prediction with different training and testing sets. As illustrated by Fig. \ref{fig:cutoff}, Our  model achieves consistent results, which further verifies the generalizability of our proposed knowledge transfer mechanism.

\section{Related Work}
\vspace{-0.1cm}
\noindent
\textbf{Knowledge Diffusion and Transfer.}
Extensive studies have been dedicated to study the diffusion of knowledge \cite{kuhn1962structure, rogers1995diffusion,  hallett2019public}, and the transfer of knowledge from science to more applicable domains like technology \cite{narin1985technology,tijssen2001global}.  The majority of these studies focus on identifying contributing factors to knowledge diffusion and transfer \cite{rossiter1993matthew,azoulay2010superstar,shi2010citing,kim2017modeling}. However, this line of work falls short in that (a) they focus primarily on successful / post-hoc knowledge diffusion and transfer, and little comparison of successful with unsuccessful transfer are presented, and (b) poorly specify what idea is being transferred because it focuses entirely at the document / invention level. In contrast, we contribute by empirically investigating properties of knowledge transfer through large-scale data analysis at the concept-level by using text mining approaches, through which we not only verified existing findings, but also revealed the significance of knowledge co-occurrence and ideational context in shaping knowledge transfer.

\para{Temporal Sequence Modelling.}
As one fundamental task in behavior modelling and NLP, numerous techniques for modelling and predicting temporal sequence have been proposed \cite{kurashima2018modeling,pierson2018modeling}. In recent years, leveraging recurrent neural network (RNN) \cite{mikolov2010recurrent} and its variants (e.g. LSTM, GRU) \cite{chung2014empirical} for sequence modelling  has been especially popular due to the structure's expressive power of temporal dynamics, and has been widely used in time series modelling \cite{lai2018modeling}.

\section{Conclusion}
In this paper, we systematically studied the process and properties of knowledge transfer from research to practice. Specifically, we used a sample of 38.6 million research papers, 4 million patents and 280 thousand clinical trials, where we leveraged AutoPhrase to extract concepts from text and focus on the applicable career of nearly 450,000 new scientific concepts that emerged from 1995-2014. Through extensive analysis, we found that `transferable' ideas distinguish themselves from `non-transferable' ideas by their (a) intrinsic properties and their temporal behavior, and (b) their relative position to other concepts. Through predictive analysis, we showed our proposed features can explain majority of transfer cases. Our research not only provides significant implications for researchers, practitioners, and government agencies as a whole, but also introduces a novel research question of real world impact for computer scientists.

\section*{Acknowledgments}

We thank Dan Jurafsky for helpful discussions and suggestions. We also thank the associate editor and reviewers for their valuable comments. Xiang Ren is supported in part by the Defense Advanced Research Projects Agency with award W911NF-19-20271 and NSF SMA 18-29268.

%Do not include this section when submitting your paper for review.

\bibliographystyle{acl_natbib}
\bibliography{refer}

\clearpage
\appendix
%\section{Appendix}
\label{sec:appendix}

\section{Details of Corpora Data}
\label{subsec:data_detail}
\textbf{Research Papers from WoS.}
The corpus covers both STEM - bio \& health sciences (16,252,065 papers), physical \& math sciences (8,390,777 papers), engineering (5,000,172 papers) and agriculture (2,568,702 papers) and non-STEM subjects - humanity (3,219,403 papers) and social science (3,146,897 papers). 
%Since WoS over-represents STEM fields (STEM fields cover ~92\% of WoS) in comparison to non-STEM fields, we sought a more balanced sample of STEM (61.18\%) and non-STEM papers (38.82\%). We chose all papers in non-STEM fields from WoS until 2016 that entail all social science (3.1M papers), humanities (3.4M papers), business (343K papers) and education papers (355K papers), totaling around 7.2M papers in total. The sample of papers in STEM fields is far larger, so except biology\footnote{We exclude biology as a STEM field because it, alone, yields 17M papers and further imbalances the sampling of fields. Our computational techniques and hardware are stretched to their max at over 18 million total texts.}, we selected all subjects including chemistry (4M papers), physics (3.1M papers), mathematics (872K papers), neuroscience (1.6M papers) and material science (1.9M papers), totaling around 11.4M papers.
The WoS dataset also includes meta-data for each paper, \ie, author name, institution, subjects, publication year and citations, which help construct measures concerning different knowledge transfer mechanisms. 

% \subsection{Patents from United States Patent and Trademark Office}
\para{Patent Documents from USPTO.}
We include 4,721,590  granted patents from the main USPTO corpus (1976-2014) covering both STEM (science, technology, engineering and mathematics) and non-STEM subjects.  %We chose all patents in non-STEM fields from USPTO until 2014 and STEM fields from USPTO until 2014, totaling around 4,721,590 patents.
Furthermore, the USPTO dataset also includes meta-data for each patent, \ie, inventor name, institution, award year and citations, which help construct several science-technology linkage measures in the knowledge transfer process. 

\para{Clinical Trials from U.S. National Library of Medicine.}
\label{subsec:data_eval}
The clinical trial dataset includes 279,195 government registered clinical trials ranging from 1900 to 2018. The corpus include the clinical trial title, their brief summary. 
%and we extract 112,389 concepts from them. 

\section{Phrase Detection Techniques, and Evaluations}
\label{subsec:eval}
The phrase detection technique we adopted is AutoPhrase~\cite{shang2018automated}, a widely-used method that extracts frequent and meaningful phrases through weak supervision. AutoPhrase first extracts single-word and multi-word expressions (\ie phrases) from the text corpus as candidate concepts, and then applies salient concept selection functions to pick the most representative concepts for each document. Given a word sequence (e.g., a sentence in an abstract), phrase segmentation can partition the word sequence into non-overlapping segments, each representing a cohesive semantic unit as illustrated in the first step in. We used default parameters as suggested by \cite{shang2018automated} in our study.

We further conducted data cleaning on the output of AutoPhrase to ensure the quality of the analyzed concepts. Specifically, we filtered out general phrases used for scientific writing (e.g. 'significantly important') and publisher name (e.g., 'Elsevier').

To quantitatively evaluate AutoPhrase for concept extraction, we randomly sampled 200 outputs and asked three experts to manually label whether they are good-quality concepts or not, where 184 (92\%) are labelled as good-quality by all three experts.

\section{Calculations of Emotionality and Accessibility}
\label{subsec:linguistic}
\textit{Emotionality} is computed as the percentage of words that were classified as either positive or negative where a concept is used. The number of positive and negative words in each article is counted by the Linguistic Inquiry and Word Count computer program (LIWC), which adopts a list of words classified as positive or negative by human readers beforehand \cite{pennebaker2015development}. We quantify accessibility through a variation of Dale Chall readability \cite{powers1958recalculation} by substituting the `easy term list' with college student vocabulary. This widely used index variable essentially measures the difficulty or appropriateness of the writing for each article. We then weighted the average Dale Chall readability score of all the documents associated with a concept.  

\section{Calculations of Graph Features}
\label{subsec:graph_features}
Given co-occurrence graph $\mathcal{G}=\{\mathcal{V},\mathcal{E}, \textbf{s}, W\}$ defined in subsection 3.2, the \textit{weighted degree} $d_i$ and \textit{weighted percentage of transferred neighbors} $p_i$ are calculated as follows.
$$d_i=\sum_{j\in\mathcal{N}_i}W_{ji}; p_i=\frac{\sum_{j\in\mathcal{N}_i,s_j=1}W_{ji}}{d_i}.$$ Different from unweighted features, weighted degree and weighted percentage use co-occurrence weights to stress the influence of high-frequency correlations. The edge weights is necessary especially when central concept co-occurs with a large amount of non-transferred concepts.   

\section{Characteristic Difference between Transferred and Non-transferred via t-test}
\label{subsec:difference}
Table \ref{table:attribute} illustrates the mean value for each attribute with regard to transferred concepts and non-transferred concepts, where we observe a statistically significant gap between the two groups. 

\begin{table}[t]
 \vspace*{-0.3cm}
\caption{Mean attribute value for transferred and non-transferred concepts. The two groups demonstrate statistically significant difference through t-test. p<0.001: ***.}
 \vspace*{-0.2cm}
 \scalebox{0.8}{
\begin{tabular}{ccc}
\toprule
Concept attribute &Transferred               &    Non-transferred            \\ 
\toprule

Adopter size ***       &     89.6   &    14.9      \\ 
Repeat usage  ***      &    10.6    &    1.2      \\ 
\midrule
Discipline diversity  ***      &   0.68     &  0.32        \\
Engineering relation  ***      &   0.15     &     0.04     \\
\midrule
Emotionality  ***      &    0.31    &     0.20     \\
Accessibility  ***     &    4.88    &    4.82      \\
\midrule
Journal Linkage  ***      &   0.28     &    0.15      \\
Univ.-Industry relation  ***      &   0.33     &     0.24     \\
\bottomrule
\end{tabular}
}
%\vspace*{1mm}
\label{table:attribute}
\end{table}

\section{Field Comparison.}
\label{subsec:field_comparison}
We studied transfer patterns in different fields. We identified the field of each concept as one of the six disciplines -- biology \& health sciences, physical \& math sciences, the humanities, engineering, agriculture, and the social sciences, based on the maximum TF-IDF value component of its field use frequency distribution. While different fields demonstrate distinct transfer rates from research to patent --- engineering 7.5\%, physical \& math sciences 1.9\%, the social sciences 1.1\%, bio \& health sciences 0.96\% (11.3\% concepts in bio \& health sciences transferred to clinical trial), agriculture 0.83\% and the humanities 0.39\% --- we found that the aforementioned features show consistent patterns in different fields.

\section{Feature Correlation}
\label{subsec:feature_correlation}
\begin{figure}[t]
 \vspace{-0.4cm}
    \centering
    \includegraphics[width=0.75\linewidth]{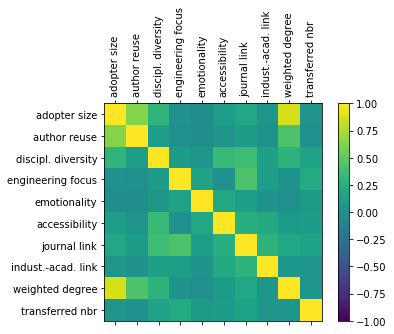}
    \vspace{-0.4cm}
    \caption{Correlations between different features. }
    \label{fig:overlap}
     \vspace{-0.3cm}
\end{figure}

We further studied the correlation between the extracted features. As illustrated in Fig.\ref{fig:overlap}, within concept individual level features, apart from hype features, and journal linkage/engineering focus, most features are rather independent. Meanwhile, graph feature `edge weight' highly correlates with hype feature. In comparison, graph feature `translated neighbor rate' brings signal not covered elsewhere, thus we conclude that modelling through both intrinsic values and graph is important.

\section{Details of Temporal Feature Model}
The RNN model is given as 
\begin{equation}
    \textbf{h}_{x,i}^{(t)} = \mathrm{RNN}\left( \textbf{h}_{x,i}^{(t-1)},\textbf{x}_i^{(t)} \right)
\end{equation}
where $\textbf{h}_x$ is the hidden states of attributes. Suppose the concept transfer status is Markovian, then the model should be 
%\begin{equation}
$$    P\left(y_i^{(t)}\middle|\;\textbf{h}_{x,i}^{(t-k)},\cdots,\textbf{h}_{x,i}^{(t-1)}\right)=
    P\left(y_i^{(t)}\middle|\;\textbf{h}_{x,i}^{(t-1)}\right)=
    g\left(\textbf{h}_{x,i}^{(t-1)}\right)$$
%\end{equation}
Here we adopt GRU as RNN and one fully connected layer with sigmoid activation as classifier $g(\cdot)$.

\section{Details on Mixed Effect Logistic Regression}
\label{subsec:glmm}
We ran a mixed effects logistic regression as a robustness check of logistic regression. Mixed effect logistic regression is a form of Generalized Linear Mixed Model (GLMM). Mixed effects logistic regression accounts for both within-concept variation (how concept use changes) and between-concept variation (how concept use differs on average), while a single measure of residual variance from the vanilla logistic regression can’t account for both.

\section{Model Training and Hyperparameters.}
\label{subsec:hyper}
To deal with the data imbalance problem -- the positive samples (concepts which will transfer in the future) are much less than the negative, we over-sample positive samples to make their amount equal to negative ones in training set while keeping the original distribution in test set. 

The hidden state size in RNN is set as $32$. We experimented on different state sizes, and $32$ achieved best performance on testing set.

%We utilize DGL~\cite{wang2019dgl} and PyTorch~\cite{paszke2017automatic} library to implement RGCN and GRU. We configure graph node embedding size and RGCN hidden states size as $8$, which performs best in grid search as there is no other downstream task to pretrain node embedding and the graph model overfits easily. The hidden state size in RNN is set as $32$. Considering both computation consumption and prediction performance, we adopt a 2-aggregate-layer aggregator for all graph models. For the remaining parameters, i.e., input history length, prediction window length and cutoff year, we vary their value and further investigate their roles in prediction.

%\section{Sensitivity Analysis}

\appendix

%\section{Appendices}
%\input{appendix.tex}

\end{document}